%% file: main.tex
\documentclass[sigconf,authorversion,nonacm]{acmart}
\makeatletter
\let\@authorsaddresses\@empty

\makeatother
\AtBeginDocument{%
  \providecommand\BibTeX{{%
    \normalfont B\kern-0.5em{\scshape i\kern-0.25em b}\kern-0.8em\TeX}}}

\usepackage[colorinlistoftodos,textsize=tiny]{todonotes}
\usepackage{multirow}
\usepackage{booktabs}
\usepackage{nicefrac}
\usepackage{amsmath,bm,dsfont,bbm,mathtools}
\usepackage{algorithmic}
\usepackage{amsthm,nccmath,lipsum,cuted}
\usepackage{xcolor}
\usepackage{colortbl}
\usepackage{graphicx}
\usepackage[lined,boxed,commentsnumbered, ruled,linesnumbered]{algorithm2e}
\usepackage{subcaption}
\usepackage{cleveref}
\usepackage{amsfonts}
\usepackage{ragged2e}

\usepackage{graphicx}
\usepackage{verbatim}
\usepackage{supertabular}
\usepackage{multicol}
\usepackage{physics}
\usepackage[footnote,draft,danish,silent,nomargin]{fixme}

\newcommand{\andcol}[1]{{\color{blue} #1}}

\graphicspath{{./figs/}}
\usepackage[utf8]{inputenc}
 \makeatletter
\newcommand*{\rom}[1]{\expandafter\@slowromancap\romannumeral #1@}
\makeatother
\title{The Case for Hierarchical Deep Learning Inference at the Network Edge}

\author{Ghina Al-Atat}
\email{ghina.alatat@imdea.org}
\affiliation{%
  \institution{IMDEA Networks Institute}
  \streetaddress{Avda. del Mar Mediterraneo, 22}
  \city{Madrid}
  \country{Spain}
  \postcode{28918}
  }

\author{Andrea Fresa}
\email{andrea.fresa@imdea.org}
\affiliation{%
  \institution{IMDEA Networks Institute}
  \streetaddress{Avda. del Mar Mediterraneo, 22}
  \city{Madrid}
  \country{Spain}
  \postcode{28918}
  }

\author{Adarsh Prasad Behera}
\email{adarshbehera.iiita@gmail.com}
\affiliation{%
  \institution{IMDEA Networks Institute}
  \streetaddress{Avda. del Mar Mediterraneo, 22}
  \city{Madrid}
  \country{Spain}
  \postcode{28918}
  }

\author{Vishnu Narayanan Moothedath}
\email{vnmo@kth.se}
\orcid{0000-0002-2739-5060}
\affiliation{%
  \institution{KTH Royal Institute of Technology}
  \streetaddress{Malvinas V{\"a}g 10}
  \city{Stockholm}
  \country{Sweden}
  \postcode{10044}
}

\author{James Gross}
\email{jamesgr@kth.se}
\orcid{0000-0001-6682-6559}
\affiliation{%
  \institution{KTH Royal Institute of Technology}
  \streetaddress{Malvinas V{\"a}g 10}
  \city{Stockholm}
  \country{Sweden}
  \postcode{10044}
}

\author{Jaya Prakash Champati}
\email{jaya.champati@imdea.org}
\orcid{0000-0002-5127-8497}
\affiliation{%
  \institution{IMDEA Networks Institute}
  \streetaddress{Avda. del Mar Mediterraneo, 22}
  \city{Madrid}
  \country{Spain}
  \postcode{28918}
  }
\begin{document}

\begin{abstract}
   % Resource-constrained Edge Devices (EDs), e.g., IoT sensors and microcontroller units, are expected to make intelligent decisions using ML algorithms, thus enabling AI at the edge of the network. Toward this end, there is a significant research effort in developing tiny ML models – DNN models with reduced computation and memory storage requirements – that can be deployed on these devices. Also, on a different front, there are research works which propose dividing large DNNs between edge devices and edge servers. In this paper, we argue for the design of ML models and algorithms tailored toward energy-efficient embedding of intelligence at the edge. In particular, neither decision-making solely on the EDs using tiny ML models nor partitioning large DNNs among EDs and ESs are not tailored for energy efficiency. Instead, we should look for design strategies where smart decisions on some data samples can be taken on EDs with minimal energy and which will eliminate these data samples from being transmitted to the ES, while data samples that require further decision-making are offloaded to the ES. We call this \textit{hierarchical inference}. We present application scenarios where hierarchical decision-making brings significant energy savings.
   Resource-constrained Edge Devices (EDs), e.g., IoT sensors and microcontroller units, are expected to make intelligent decisions using Deep Learning (DL) inference at the edge of the network. Toward this end, there is a significant research effort in developing tinyML models – Deep Learning (DL) models with reduced computation and memory storage requirements – that can be embedded on these devices. However, tinyML models have lower inference accuracy. %On a different front, DNN-partitioning and EarlyExit techniques were studied for running large-size DL models on mobile devices such as smartphones. Inference offloading is another research direction that does load balancing for distributed DL inference between EDs and Edge Servers (ESs). 
   On a different front, DNN partitioning and inference offloading techniques were studied for distributed DL inference between EDs and Edge Servers (ESs).
   In this paper, we explore Hierarchical Inference (HI), a novel approach proposed in~\cite{moothedath2023} for performing distributed DL inference at the edge. Under HI, for each data sample, an ED first uses a local algorithm (e.g., a tinyML model) for inference. Depending on the application, if the inference provided by the local algorithm is incorrect or further assistance is required from large DL models on edge or cloud, only then the ED offloads the data sample. At the outset, HI seems infeasible as the ED, in general, cannot know if the local inference is sufficient or not. Nevertheless, we present the feasibility of implementing HI for machine fault detection and image classification applications. We demonstrate its benefits using quantitative analysis and argue that using HI will result in low latency, bandwidth savings, and energy savings in edge AI systems.
\end{abstract}
\maketitle
\input{intro}
\input{related}

\input{rollingelement}
\input{CIFAR-10}

\input{dogbreed}
\input{comparison}

\input{conclusion}
\input{appendix}
\bibliographystyle{ACM-Reference-Format}
\bibliography{main}

%\bibliographystyle{IEEEtran}
%\bibliography{main}
\end{document}

%% file: intro.tex
\section{Introduction}
Deep Learning (DL) models are compute and memory intensive and have been traditionally deployed in the cloud.  
However, recently, an increasing number of applications in Cyber-Physical Systems (CPS), remote healthcare, smart buildings, intelligent transport, etc. use DL inference at the edge of the network. 
%Monitoring a process/phenomenon of specific interest at the network edge is prevalent in Cyber-Physical Systems (CPS), remote healthcare, smart buildings, intelligent transport, etc., which are essential building blocks of smart cities. 
%Today’s monitoring systems extensively use Internet-of-Things (IoT) sensors and an increasing number of monitoring applications at the network edge are using Machine Learning (ML) inference, in particular Deep Neural Network (DNN) inference, for detecting essential events. 
This initiated a major research thrust in developing small-size ML models (S-ML) – ML models with reduced computation and memory storage requirements – and deploying them on resource-constrained Edge Devices (EDs) such as IoT sensors, wearable devices, mobile phones, drones, and robots. Such deployments are enabled by advancements in hardware and the rapid evolution of model compression techniques \cite{Deng2020,Sanchez2020}.

Research works on DL inference at the edge can be broadly classified into 1) tinyML, 2) DNN-partitioning, and 3) inference offloading. TinyML research focuses on enabling extremely resource-limited IoT devices such as micro-controller units (MCUs) to perform on-device DL inference using custom-designed S-ML models. %The benefits of using tinyML models for on-device inference include, 1) improved responsiveness, as it circumvents communication time and remote inference time, 2) bandwidth savings, and 3) may result in energy savings if the energy consumption of S-ML for providing inference for a sample is less than the transmission energy of the sample. 
 Doing inference at the ED using S-ML saves network bandwidth, improves responsiveness and energy efficiency of the system~\cite{Zhang2017, Fedorov2019}. Recent advances in tinyML research enable sophisticated applications performing tasks on images and audio that go far beyond the prototypical IoT applications that include monitoring environmental conditions (e.g., temperature, carbon dioxide levels, etc.), machine vibrations for predictive maintenance, or visual tasks such as detecting people or animals~\cite{Njor2022}. However, tinyML models have relatively poor inference accuracy due to their small size, which limits their generalization capability and robustness to noise.

The authors in~\cite{Kang2017} proposed DNN partitioning that divides the inference execution of large-size DNNs between an ED and an Edge Server (ES). However, the benefits of this technique were only realized for computationally powerful EDs (such as smartphones) with mobile GPUs \cite{jointoptimization,combiningdnn}. Finally, works on inference/computation offloading propose load-balancing techniques that divide the inference computational load between the ED and the ES \cite{Wang2019,Ogden2020,Nikoloska2021,Fresa2022}.

\begin{figure} [ht] 
    \begin{center}
  \includegraphics[width=7.5 cm]{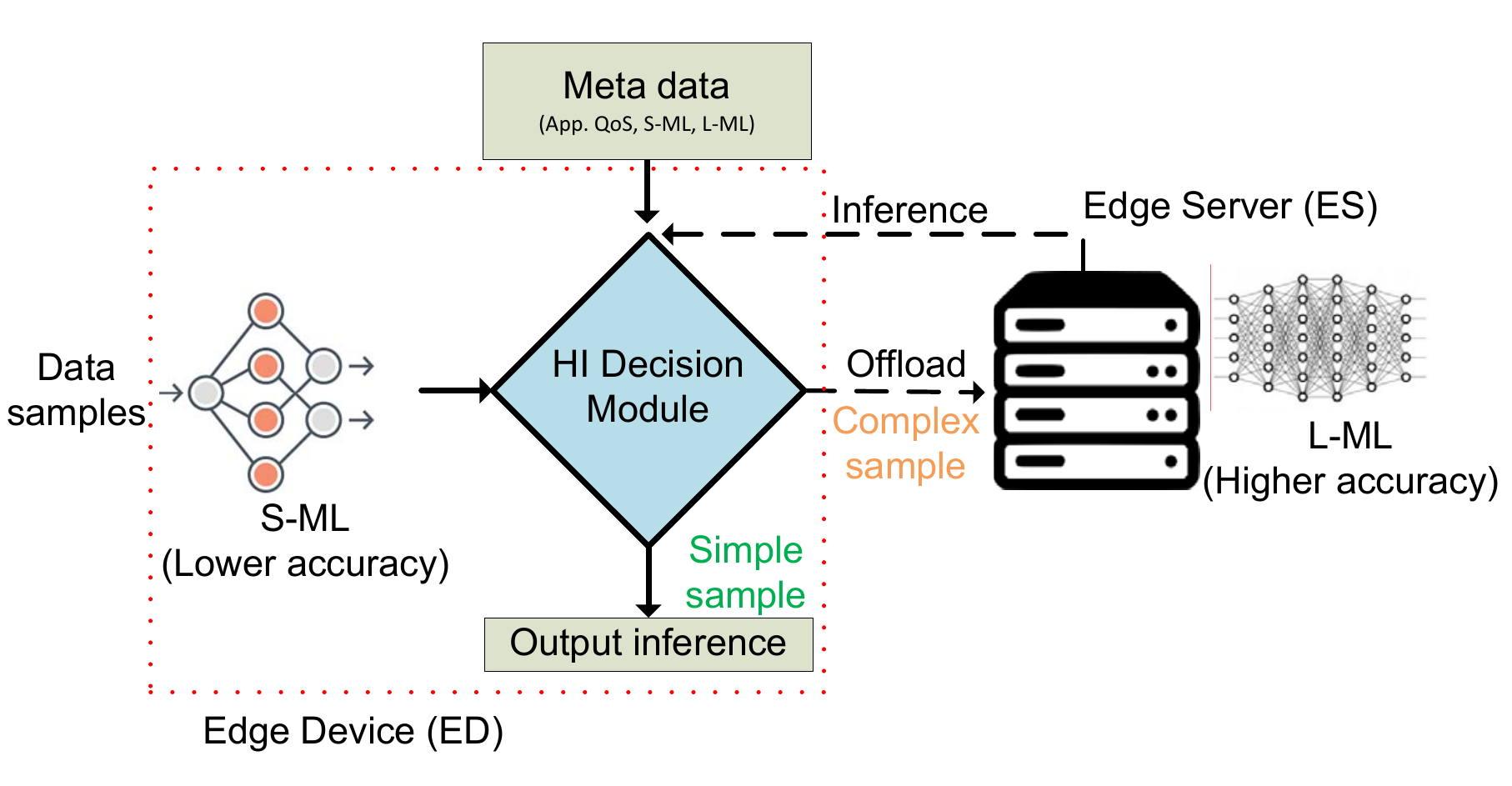}
    \end{center} \caption{Schematic of HI framework.} \label{fig:HI-system}
\end{figure}

\begin{figure*}[h]
\centering
\begin{subfigure}{0.2\textwidth}
  \centering
  \includegraphics[width=\linewidth]{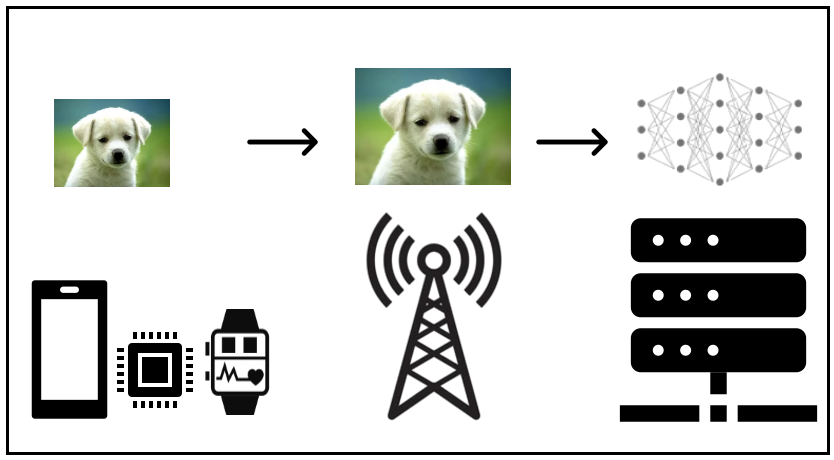}
  \caption{Full Offload}
  \label{fig:subfigure1}
\end{subfigure}%
\begin{subfigure}{0.2\textwidth}
  \centering
  \includegraphics[width=\linewidth]{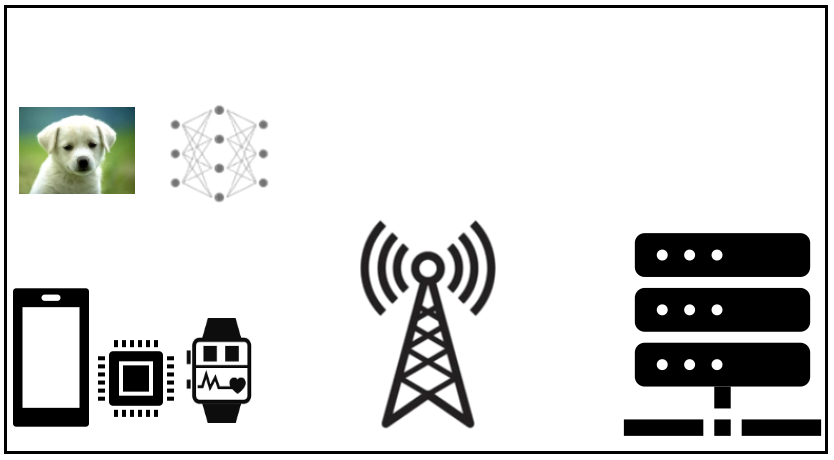}
  \caption{TinyML }
  \label{fig:subfigure3}
\end{subfigure}%
\begin{subfigure}{0.2\textwidth}
  \centering
  \includegraphics[width=\linewidth]{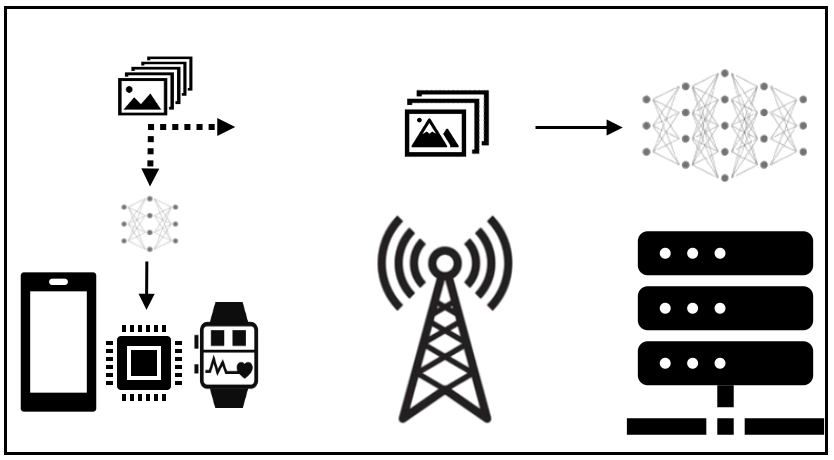}
  \caption{Inference Offloading}
  \label{fig:subfigure2}
\end{subfigure}%
\begin{subfigure}{0.2\textwidth}
  \centering
  \includegraphics[width=\linewidth]{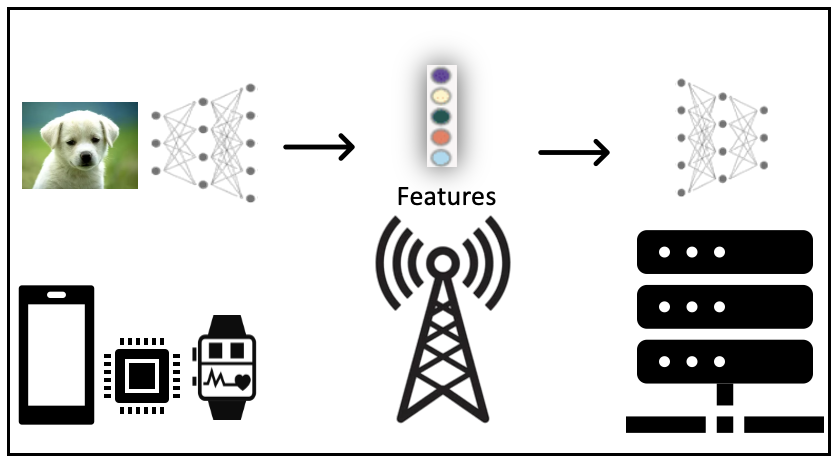}
  \caption{DNN-partitioning }
  \label{fig:subfigure4}
\end{subfigure}%
\begin{subfigure}{0.2\textwidth}
  \centering
  \includegraphics[width=\linewidth]{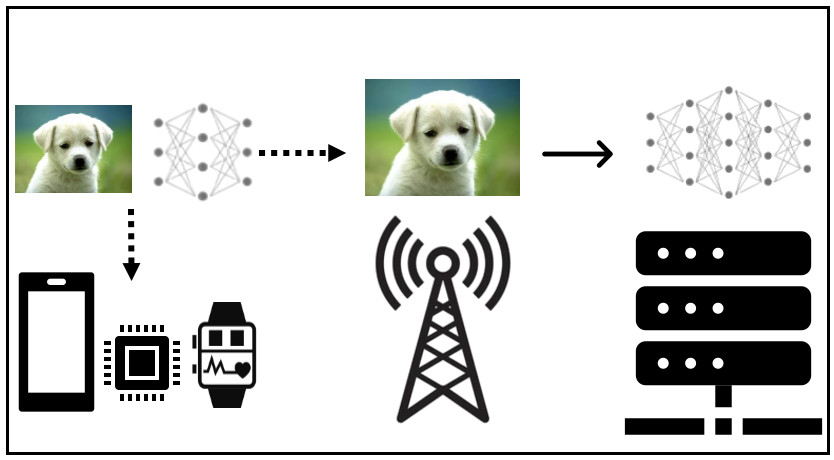}
  \caption{Hierarchical Inference}
  \label{fig:subfigure5}
\end{subfigure}
\caption{Different approaches for  DL inference at the Network Edge}
\label{fig:DL_Approaches}
\end{figure*}

In contrast to the above works, we propose \textit{Hierarchical Inference}, a novel framework for performing distributed DL inference at the edge. Consider the system where the ED is embedded with an S-ML model and enlists the help of ES(s) or cloud on which a stat-of-the-art large-size ML model (L-ML) is available. HI differentiates data samples based on the inference provided by S-ML. A data sample is \textit{simple data sample} if S-ML inference is sufficient, else it is a \textit{complex data sample} requiring L-ML inference. The key idea of HI is that only complex data samples should be offloaded to the ES or cloud. Figure~\ref{fig:HI-system} shows the HI framework. The HI decision module takes as input the S-ML inference and uses the metadata about S-ML, L-ML, and the application QoS requirements to decide whether a sample is complex or simple and accordingly offload it or not. 

Figure~\ref{fig:DL_Approaches} illustrate different approaches for DL inference at the edge. In contrast to tinyML research, HI considers inference offloading. Further, HI examines the S-ML inference before making an offloading decision which is in stark contrast to existing inference offloading algorithms.  Unlike DNN partitioning, HI is appealing for resource-constrained devices as one may design customized S-ML models for these devices and use off-the-shelf L-ML models on ESs or the cloud. 

\subsubsection*{Advantages of HI} On the one hand, performing inference on EDs saves network bandwidth, improves responsiveness (reduces latency), and increases energy efficiency. On the other hand, doing inference on ESs results in high accuracy inference. HI reaps the benefits of performing inference on EDs without compromising on the accuracy that can be achieved at ESs. To see this, note that HI offloads only complex data samples which need further inference assistance and will (most likely) receive correct inference from L-ML at the ES, while simple data samples which are either redundant or receive correct inference from S-ML are not offloaded, thus saving bandwidth, reducing latency and energy for transmission and computation on L-ML. Although we use the term `S-ML', it need not be an ML algorithm but can be any signal processing or statistical algorithm. 

Note that, under HI, performing additional inference locally on all images incurs an extra cost. However, in scenarios where the cost of local computation is lower than that of transmission, this approach would eventually result in significant cost savings, especially since such scenarios are frequent.
    %Nevertheless, there is an ongoing development of autonomous cars that connect over 5G. Connectivity to the network will in requires real-time inference and most existing autonomous cars use DL inference from local computing. Nevertheless, there is an ongoing development of autonomous cars that connect over 5G. Connectivity to the network will increase the safety of the cars as the DL models on the cars may not be able to provide correct inference to unknown situations that are not part of the training dataset. Thus, communication with the infrastructure through V2I and with the nearby vehicles using V2V protocol will indeed .  Responsiveness (low latency) is a key requirement for autonomous cars. It is imperative that real-time decisions making on these cars should be made locally  
%Bandwidth Savings: Machine fault detection
%Energy savings: Battery-limited IoT sensors

\subsubsection*{Challenges of HI} There are a few challenges to implementing HI. 
\begin{enumerate}
\item Differentiating simple and complex data samples is the key to HI. However, in general, it is hard to do this differentiation as the ED does not know a priori if the inference output by S-ML is the ground truth or not.
%or whether the sample is important or redundant.
\item There are certain requirements for S-ML in order to enable HI. The S-ML should be small enough in size such that, 1) it should be viable for embedding on ED, 2) the computation energy required for S-ML inference should be less than the transmission energy required for transmitting a data sample, and 3) the S-ML should have reasonable inference accuracy such that for the application of interest the fraction of simple data samples should be higher than that of complex data samples.
\end{enumerate}
%\todo[inline]{Should we rephrase the second challenge? (1) is a prerequisite for S-ML but not a challenge of HI. (2) Is correct in some particular setting, but as we are currently talking about HI in general, there might be other cases where we would like to offload some samples even with a large communication cost; for example when the SML resources are strictly limited. }
\begin{comment}
as shown in Fig. \ref{fig:partition}.
\begin{figure} [ht] 
    \begin{center}
  \includegraphics[width=7.5 cm]{Uploads/partition.png}
    \end{center} \caption{Partition of the data samples that is decided based on the S-ML output.} \label{fig:partition}
\end{figure}
\end{comment}

%Despite the above challenges, HI is feasible for several applications. 
%Autonomous driving is a natural example where HI is clearly feasible and is certainly useful. Note that unknown situations or unknown encounters or bad weather are the key challenges for autonomous driving. To overcome this, an autonomous vehicle can use DL inference from local computing when the inference accuracy is high for the data collected (simple data samples) but when it encounters unknown situations it may offload the data (complex data samples) to the edge if the local decisions have low inference accuracy. 
Despite the above challenges, in this work, we demonstrate its feasibility by implementing it in three application scenarios: 1) rolling element fault diagnosis, 2) CIFAR-$10$ image classification, and 3) dog breed image classification. The first scenario is a specific example of machine fault detection, where the S-ML is a simple threshold rule on the statistical average of the vibration data. For image classification, we use TFLite and design a customized S-ML for performing inference on images. For each use case, we also provide a quantitative analysis of the benefits of HI. Finally, we also compare the accuracy and delay performance of HI with existing techniques for DL inference: 1) 
 tinyML (no offload), 2) DNN-partitioning, 3) Offloading for Minimizing Delay (OMD), and 4) Offloading for Maximizing Accuracy (OMA). 

%7 key requirements for Trustworthy AI by EU. The requirement on Societal and environmental well-being laid by the EU for AI algorithms refers to the carbon footprint of these algorithms both during the training as well as inference stage of the ML algorithms.

%tinyML - On-device machine learning applications in the single mW and below.

%embed IoT sensors with low-complex algorithms that can provide inference on data samples that are easier to discern and offload the data samples which require complex algorithms such as DNNs to discern. We use a couple of use cases from the anomaly detection and demonstrate that diving the decisions using the above approach could lead to unprecedented energy gains in the system without compromising the battery lifetimes of the IoT sensors. 

%% file: related.tex
\section{Related Works}
Since the advent of AlexNet \cite{AlexNet} a decade ago, there has been an explosion in research on bigger DL models with an increasing number of layers and nodes per layer resulting in models with billions of parameters. Examples include the state-of-the-art image classification model BASIC-L (with 2.4 billion parameters)~\cite{chen2023} and the revolutionary language model chat-GPT (with 175 billion parameters)~\cite{chatgpt}. However, large DL models are often over-parameterized and there has been significant research on model compression techniques that reduce the model size by trading accuracy for efficiency, e.g., see~\cite{Song2016, He2018}. Efficient model compression techniques have resulted in mobile-size DL models including SqueezeNets \cite{Squeezenet}, MobileNets \cite{Mobilenet, Sandler2018}, ShuffleNet \cite{ShuffleNet}, and EfficientNet \cite{Tan2019}. The efforts for designing small-size DL models were further fueled by the need to perform DL inference on edge devices, which span moderately powerful mobile devices to extremely resource-limited MCUs. In the following, we discussed the related work on DL inference at the edge.

\subsubsection*{TinyML} The tinyML research~\cite{Sanchez2020} focuses on embedded/on-device ML inference using small-size ML models on extremely resource-constrained EDs such as IoT sensors including micro-controller units (MCUs). The motivation for tinyML stems from the following drawbacks of sending data over a network  \cite{Njor2022}: 1) security and privacy of data, 2) unreliable network connection, 3) communication latency, and 4) significant transmission energy consumption. A suite of tinyML models is now available for MCUs enabling a wide variety of complex inference tasks such as image classification, visual wake word, keyword spotting \cite{Banbury2021}, monitoring underwater water environments\cite{Adib_2022}, etc.
However, tinyML models have poor inference accuracy (relative to large-size DL models) due to their small size, which limits their generalization capability and robustness to noise. Further, once a tinyML model is trained (on an edge server/cloud) and deployed on an IoT sensor or an MCU, improving its accuracy using active learning on each data sample may not be possible due to the resource constraints of the device. Therefore, to improve inference accuracy the EDs need to offload the data samples by enlisting the help of an Edge Server (ES) or a cloud, where L-ML models are deployed. 

\subsubsection*{DNN Partitioning} For the scenarios where the ED is a powerful mobile device, such as a smartphone, the authors in~\cite{Kang2017} proposed DNN partitioning, where  the front layers of the DNN are deployed on mobile devices while the deep layers are deployed on Edge Servers (ESs) to optimize the communication time or the energy consumption of the mobile device. The premise of this technique is that the data that needs to be transmitted between some intermediate layers of a DNN is much smaller than the initial layers. Following this idea, significant research work has been done that includes DNN partitioning for more general DNN structures under different network settings \cite{Chuang2019,Li2020} and using heterogeneous EDs \cite{Hu2022}, among others. However, as we will show later, for DNN partitioning to be beneficial, the processing times of the DNN layers on the mobile device should be small relative to the communication time of the data generated between layers. Thus, works on DNN partitioning (e.g., see \cite{jointoptimization,combiningdnn}) require mobile GPUs making this technique infeasible for extremely resource-constrained IoT sensors and MCUs. 

\subsubsection*{Inference Offloading} Inference offloading or computation offloading is a load-balancing technique for partitioning the set of data samples between the ED and the ES for computing the inference. It is worth noting that computation offloading between EDs and ESs was extensively studied in the literature, e.g., see~\cite{Mach2017,Feng2022}. However, the majority of the works studied offloading generic computation jobs, and the aspect of accuracy, which is relevant to offloading data samples, has only been studied recently~\cite{Wang2019,Ogden2020,Nikoloska2021,Fresa2022}. These works compute offloading decisions using one or more of the following quantities:  job execution times, communication times, energy consumption, and average test accuracy of the ML models. Note, however, that the average test accuracy may not be the right indicator for the correct or incorrect inferences provided by an S-ML across different data samples and thus it may result in adversarial cases where most of the data samples that are scheduled on S-ML may receive incorrect inference. In contrast to inference offloading HI first examines the S-ML output to determine if the data sample is simple or complex and only offloads if it is a complex data sample. We note that early exiting within the layers of DNNs, proposed in \cite{teerapittayanon2016}, received considerable attention recently. This technique could be used in conjunction with the above DL inference approaches, including HI, to further reduce the latency in reference.

In our previous work~\cite{moothedath2023}, we proposed the idea of HI and focused on a specific technique for realizing it for an image classification application by using an existing S-ML model. In contrast, in this paper, we provide a general definition of HI, consider multiple use cases (machine fault detection and dog breed image classification), design S-ML algorithms, and provide a quantitative comparison with existing DL inference at the edge.

%% file: rollingelement.tex
\section{HI for Rolling Element Fault Diagnosis}
\begin{figure} [ht] 
    \begin{center}
  \includegraphics[width=7.5 cm]{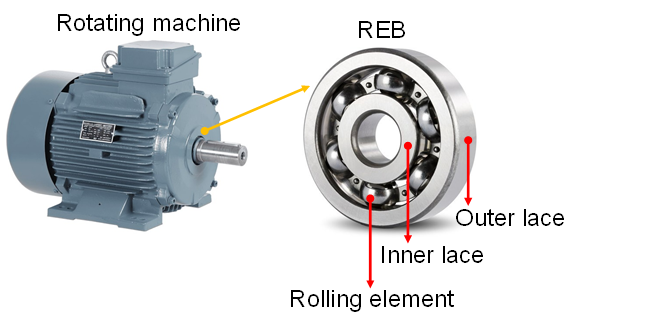}
    \end{center} \caption{An REB at the driver end of a rotating machine.} \label{REB}
\end{figure}

%Machine fault detection is a classical problem and a great deal of research has been done on diagnosing machine faults~\cite{}. 
In order to demonstrate how HI can be used in machine fault detection we turn to the motor bearing dataset, provided by the Case Western Reserve University (CWRU) bearing data centre, a standard vibration-based dataset used in Rolling Element Bearings (REBs) fault diagnosis~\cite{CWRUDataset,Wade2015}. The failure of REBs is one of the most frequent reasons for the breakdown of rotating machines.  
In Figure~\ref{REB}, we show an REB that is located at the driver end of the rotating machine. At the other end of it lies the fan end REB. A faulty REB may have a fault on the inner lace or outer lace or in the rolling element. 
%Rolling Element Bearings (REBs) are the most prevalent components of rotating machines, and their failure is on of the most frequent reason for machine breakdown.

The CWRU dataset consists of $161$ datasets, each collected for driver-end REB and fan-end REB under different settings, which includes samples collected at $12$ kHz and $48$ kHz frequencies under varying motor loads ($0-3$ hp), varying shaft speeds ($1730-1797$ rpm), and for three fault widths $0.18$, $0.36$, and $0.54$ mm. For a given motor load and shaft speed, the task is to identify $10$ states of the rotating machine, which include a normal functioning state and nine fault states resulting from three fault widths seeded on each inner lace, outer lace, and ball bearing.   

A state-of-the-art CNN proposed in \cite{Wen2018} identifies each of these states with $99.6$\% accuracy. The CNN is trained on $64 \times 64$ grey images created from batches of $4096$ consecutive samples. The CNN consists of $8$ layers and it cannot be deployed on resource-constrained sensors to do local inference. Instead, it needs to be deployed on an ES or in the cloud and transmit all the data from the sensors. Consider as an example that a factory floor consists of $100$ rotating machines. Since each rotating machine contains more than two REBs, the bandwidth required to transmit all the data to an ES or cloud (on which the CNN is deployed) in order to monitor the state of all REB is at least $76.8$ Mbps, assuming that the sampling frequency is $48$ kHz and each sample value is stored in a $2$ byte register. The question we ask is: can we achieve $99.6$\% accuracy in classifying the states of the REBs without needing to use such high bandwidth?

\begin{figure} [ht] 
    \begin{center}
  \includegraphics[width=9 cm]{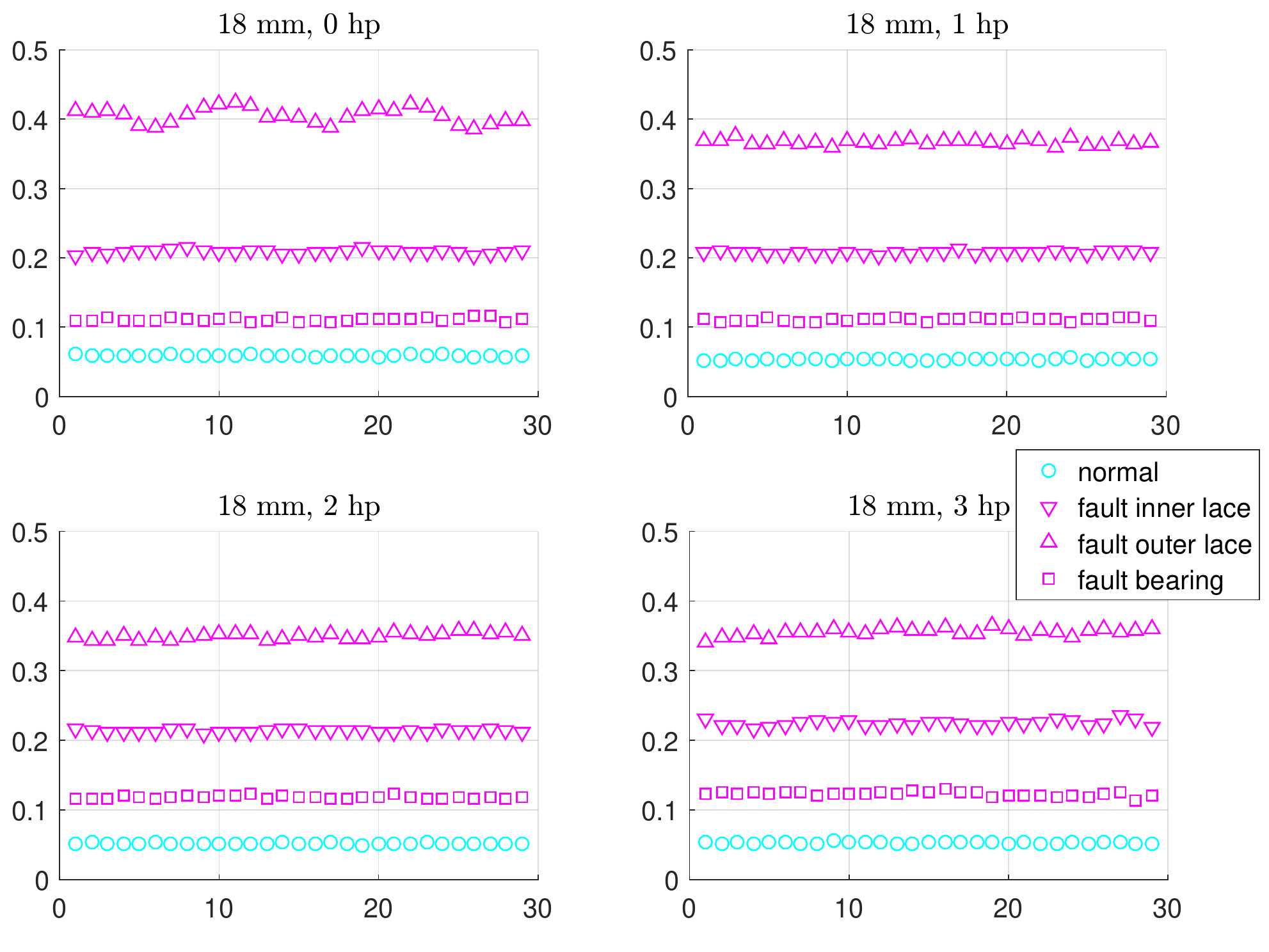}
    \end{center} \caption{Time series vibration data averaged over every $4096$ samples for aprox. 120,000 samples collected at $18$ mm fault width.} \label{fig:avg4096_18mm}
\end{figure}

\begin{figure} [ht] 
    \begin{center}
  \includegraphics[width=9 cm]{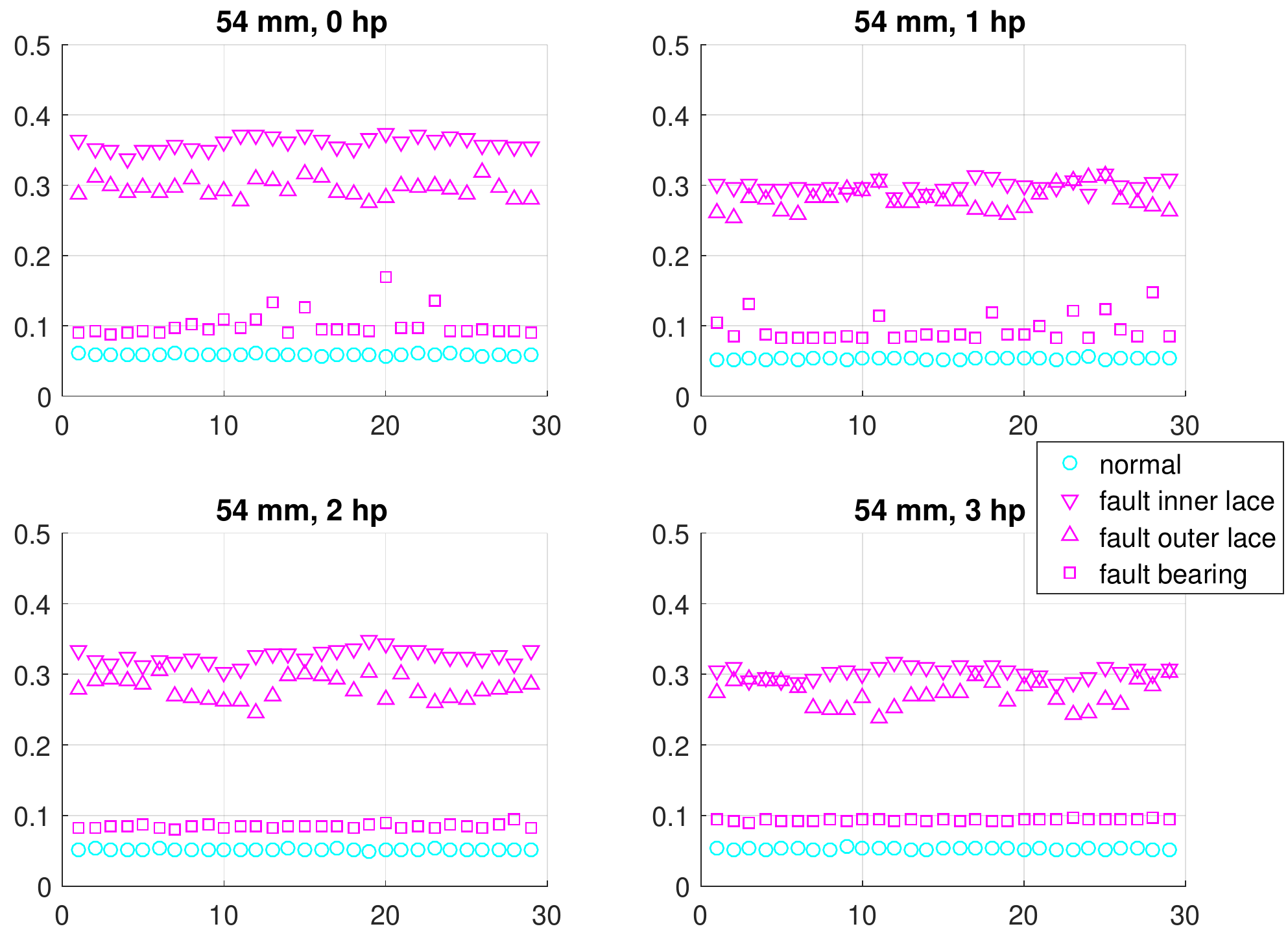}
    \end{center} \caption{Time series vibration data averaged over every $4096$ samples for aprox. 120,000 samples collected at $54$ mm fault width.} \label{fig:avg4096_54mm}
\end{figure}

In order to answer the above question, we make a simple observation that the REBs (or in general machines) work in a normal state for hundreds (or thousands) of hours. If we can differentiate normal state from \textit{not} normal state (any of the faulty states) on the sensor, then we save almost the entire bandwidth by only offloading the samples to CNN when not in the normal state. To this end, we analyzed the CWRU dataset and computed the average for batches of $4096$ consecutive vibration samples with series lengths of approximately $120,000$ from different datasets. Note that to compute the average values a sensor doesn't need to store all the $4096$ samples. Instead, it can use a simple moving average. In Figures \ref{fig:avg4096_18mm} and \ref{fig:avg4096_54mm}, we plot the average values under normal state (in cyan) and under faulty states (in magenta) for different motor loads at $18$ mm and $54$ mm fault widths, respectively. For the dataset in Figure \ref{fig:avg4096_18mm}, it is easy to verify that by setting a threshold $0.07$ below which the REB is in normal state and not in normal state, otherwise, provides $100$\% accuracy for normal vs not normal classification. Note that this simple threshold rule is the S-ML in our definition of HI. Further, for this scenario, all the states can be classified with $100$\% accuracy as the data points are separable via thresholds. 

In Figure \ref{fig:avg4096_54mm}, however, all the states are not separable. In particular, the inner lace and outer lace faults cannot be differentiated using thresholds. Nevertheless, normal and not normal states can still be differentiated using a threshold of $0.07$. Thus, the sensor only transmits data samples that correspond to one of the fault states, which of course can then be classified by the CNN at ES or cloud. 
Thus, from the above data analysis, we conclude that a sensor can make a local decision about normal state or not normal state of the REB using a basic computation of averages. Since an REB is in normal state for longer periods, significant bandwidth savings can be obtained using HI. It is worth noting that, here the samples collected in the normal state are indeed simple data samples and the samples collected in a faulty state are complex data samples. Furthermore, since computing a moving average is not compute-intensive, the sensor will save all the transmission energy that would have been spent in transmitting the simple data samples. Thus, using HI results in potential energy savings for IoT devices that in turn improves their battery lifetimes.

%We use the subset of data which is collected at fault widths seeded on the inner lace, outer lace, and rolling element using electro-discharge machining. 
%We show that the normal operating condition of the machine can be easily identified in contrast to the fault conditions. The dataset we use is the most widely used motor bearing data provided by Case Western Reserve University~\cite{Wade2015}. The dataset consists of time series data obtained using an accelerometer. There are four states: 1) normal operation, 2) rolling element fault, 3) inner race fault, and 4) outer race fault. The data was collected under different load conditions, $0$ hp (no load), $1$ hp, $2$ hp, and $3$ hp. For each state and load condition data samples are taken at $48K$ samples per second and the data is collected for approximately $5$ seconds. We provide the sample mean and standard deviation for the time series data in the following table.

%% file: CIFAR-10.tex
\section{HI for CIFAR-10 Image Classification 
} \label{usecase2} We choose Image classification as our second use case as it is a fundamental task %that attempts to recognize the image as a whole and assign a label to it. Several research fields, such as 
inherent to a variety of applications including object detection, image analysis, video analytics, and remote sensing. %rely heavily on image classification as a critical step.
Significant efforts have been devoted to developing advanced classification approaches and techniques aiming to improve classification accuracy (e.g., see \cite{Gong1992}, \cite{foody1996}, \cite{Gallego2004}). 
We choose \textit{CIFAR-$10$} image dataset consisting of $50,000$ training samples, $10,000$ test samples and $10$ classes. We design and train an S-ML that can be embedded on a resource-constrained ED such as MCU with eflash memory size of order $1$ MB and SRAM of order few hundred KB. The details of the ML models are given below.
\begin{itemize}
    %\item \textbf{CIFAR-$\mathbf{10}$ Dataset:} The CIFAR-$10$ dataset consists of $60,000$ colored images with a total of $10$ classes
    %: 'airplane', 'automobile', 'bird', 'cat', 'deer', 'dog', 'frog', 'horse', 'ship', and 'truck' 
    %\cite{krizhevsky2009}.  Each class has $6,000$ images. The dataset consists of $50,000$ training samples and $10,000$ test samples. %Each image is of $32 \times 32 \times 3$ color (RGB) pixels and the value of a given pixel in each red, green and blue component ranges between $0$ and $255$.
    \item \textbf{S-ML:} We build a convolutional neural network (CNN) with five layers: a convolutional layer, a max-pooling layer, a flatten layer, and two fully-connected dense layers. We use tensorFlow's Keras API to train it on the CIFAR-$10$ dataset. 
    %We use a typical architecture for image classification tasks, that has been shown to perform well on a variety of datasets. 
    Our final quantized TFLite mode has an inference accuracy of $62.58\% $ and is $0.45$ MB, a size suitable for deploying on MCUs.
    \item  \textbf{L-ML:} %The ES has more flexibility in terms of the ML models it can deploy due to the relaxation of constraints on computing power, memory capacity, and storage space. It is capable of handling large and complex ML models, enabling it to host any state-of-the-art large-scale ML model with high accuracy.
    Many existing state-of-the-art DNN models have already demonstrated top-1 accuracy on the CIFAR-$10$ dataset, in particular, with some models achieving up to $99.5\%$ \cite{ViT-H/14}. For this work, we choose EfficeintNet \cite{ruseckas_efficientnet_cifar10}, which achieves an accuracy of $95\%$.
    % \item \textbf{Performance Measures}: For image classification tasks, the goal is to accurately predict the correct class for a given image. Therefore, accuracy in this context stands for the portion of falsely classified images. Further to this, we add the dimension of cost. Our ideal approach would be one that optimize for cost and accuracy. 
%First, we load and preprocess the data. We download the CIFAR-$10$ dataset and preprocess it by scaling pixel values and one-hot encoding the class labels. Next, we define our machine learning model. 
\end{itemize}
 %TensorFlow Lite (TFLite) models are well-suited for small machine learning (S-ML) applications due to their smaller size and faster processing compared to full TensorFlow models. This makes them ideal for deployment on resource-constrained devices. 
Exploring the idea of HI, \textit{simple data samples} in this context would be the images that the S-ML embedded on the IoT device is able to classify correctly, otherwise they are considered as \textit{complex data samples}. The data sample will initially undergo inference by the S-ML. The ED will decide to either accept the inference or consider it incorrect and offload the data sample to the L-ML deployed on the ES.  \\
For the ED to make this decision, we use two key ideas proposed in~\cite{moothedath2023}. The first idea is with regard to how to quantify the  confidence of the S-ML's inference. For image classification, a DNN outputs a probability mass function (pmf) over the classes. Note that one can obtain such pmf for other ML classification algorithms (such as linear classifier) by normalizing the output with the sum of the values output for each class. We use the maximum probability value, denoted by $p$, from the pmf as the confidence of S-ML. This is justified because it is a standard approach to declare the class corresponding to the maximum probability $p$ as the true class. 
%Any appropriate metric can be utilized to examine or track the validity of the S-ML inference. In fact, t
%A data sample is \textit{simple data sample} if the inference provided by S-ML on that data sample is correct, otherwise, it is a \textit{complex data sample}. HI mandates that only complex data samples should be offloaded to ES. 
%The S-ML uses $p$ (\textit{the maximum probability in the probability distribution over the classes output by the S-ML}) for achieving top-1 accuracy and making predictions to assign labels to images. 
The ED will utilize the same metric to examine or track the validity of the S-ML inference and decide whether to accept the inference or delegate it to the L-ML that resides on the ES. This means that the local inference of data sample $i$ will be accepted or rejected based on its associated $p_i$. The ED's decision rule for data sample $i$, denoted by $\delta (i)$, will be based on comparing $p_i$ with a threshold $\theta\in[0,1)$:
\[
    \delta(i) = \left\{\begin{array}{lr}
        \text{Offload} & \text{if } p_i< \theta  \\
        \text{Do not offload} & \text{if }  p_i \geq \theta 
        \end{array}\right. 
  \] %The second idea we borrow from \cite{moothedath2023} is the cost model. If the S-ML's inference is rejected and the data sample $i$ is offloaded to the ES, a fixed cost $0 \leq \beta < 1$ will be received for offloading. This offloading cost, $\beta$, may in practice be connected to transmission energy or remote inference delay. 
The second idea adopted from \cite{moothedath2023} is the cost model. If the S-ML's inference is rejected and the data sample $i$ is offloaded to the ES, a fixed cost $0 \leq \beta < 1$ will be received for offloading. This offloading cost, $\beta$, can be adjusted to consider different factors that may be specific to the context of the task and the user, rather than being limited to a fixed cost tied to a specific metric. In practice, it may  correspond to transmission energy consumption or data transfer cost or remote inference delay. We choose to introduce HI using an abstract cost for offloading to enhance flexibility in the decision-making process and make it adaptable to diverse scenarios and requirements. This abstract cost may differ across various setups, allowing our decision algorithm to make more informed decisions that balance various costs and benefits, such as network conditions, resource availability, battery life, and user's personal preferences. Furthermore, denote by $\eta_i$ the additional cost obtained from the ES inference of the data sample $i$. It is equal to $1$ if the L-ML's inference is incorrect and is equal to $0$, otherwise. Alternatively, if data sample $i$'s accepted local inference is correct, then the incurred cost, denoted by $\gamma_i$, is $0$; otherwise, it is $1$. Thus, the cost for data sample $i$ using HI is given by:
  %\todo[inline]{Do we need $\eta_i$ here? Technically it is okay, but then we have to add it to both cases. We are incurring local inference cost irrespective of the decision (?), and then the cost of correct local inference $\neq0$}
  \[
    \mathcal{C}_i = \left\{\begin{array}{lr}
        \beta + \eta_i & \text{if } p_i< \theta  \\
        \gamma _i & \text{if }  p_i \geq \theta  
        \end{array}\right. 
\]\begin{figure} [ht] 
    \begin{center}
  \includegraphics[width=6 cm]{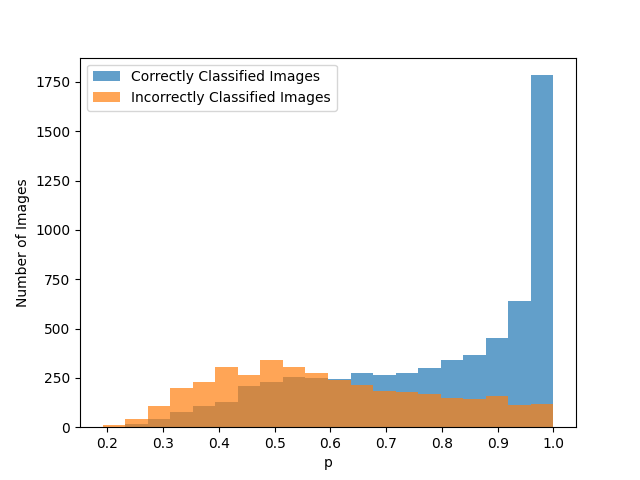}
    \end{center} \caption{Classification of CIFAR-10 by the S-ML model.} \label{pCifar}
\end{figure} 

Figure \ref{pCifar} shows the number of correctly and incorrectly classified images of the CIFAR-$10$ dataset using our S-ML model as a function of $p$.
%"(the maximum probability in the probability distribution over the classes output by the model)"
The figure clearly illustrates that for $p$ greater than $0.6$, the number of correctly classified images becomes  higher than the incorrectly classified ones. Thus, it serves as an appropriate threshold candidate to decide whether to accept the local inference or not. In fact, brute-force search shows that the optimal probability threshold $\theta^*$ that yields the minimum cost for CIFAR-10 dataset is $0.607$.

%\todo[inline]{Need to define optimum. If it is minimizing cost, with what $\beta$ we are obtaining the optimum? This optimum is independent of the intersection point in the graph(Even for $\beta=0.5$ (?), which is a mistake present in the HI paper currently under Sigmetrics revision)}
 \textsc{HI Approach}: The data samples will be subjected to inference using the S-ML embedded on the ED. The ED will decide to offload only the data whose inference's associated $p$ is less that the optimal threshold $\theta^* = 0.607$. 
  \subsubsection*{Results and Discussion} Implementing the complete offload approach, which involves running a state-of-the-art DNN on the Edge Server and offloading the whole data $(10000$ images$)$, the ES achieves an accuracy of $95\%$. Only $500$ out of $10000$ images were falsely classified, which is considered a top-tier accuracy. However, it comes at a cost of $10000 \, \beta +500$, meaning that this approach is expensive. Conversely, the complete local approach, which involves accepting all of the local inferences produced by the S-ML embedded in the ED, can reduce the cost to $3742$, which is a significant reduction particularly when $\beta$ is large and close to one. However, it entails a significant trade-off in terms of inference accuracy $62.58\%$, which is insufficient to ensure reliability in the majority of use cases. 

Following the HI approach  with $\theta^* = 0.607$, 
    only $35.5\%$ of the data is offloaded $(3550$ images$)$ of which $71$ images of them were misclassified by the ES. This yields a cost of $(3550 \beta + 71)$. Moreover, out of the $6450$ accepted local inferences, $1577$ were misclassified (which contribute an extra $1577$ to the cost). This means that the total cost of HI is $3550 \beta + 1648$. Since $1648$ images are misclassified under HI, 
    it has $83.52\%$ accuracy.  HI can result in a cost reduction of $ \big( \frac{6450 \, \beta - 1148}{10000 \, \beta +500} \times 100 \big) \%$ compared to offloading all of the images. Thus, HI can provide the best of both worlds, namely significant cost reduction while still achieving a dependable level of accuracy based on the cost $\beta$. 
    %Table \ref{table:comparisonCifar} below compares our method with the two baselines (full offload and no offload). 
    %Note that $\beta$ is a design parameter that reflects the designer's preference for balancing the costs of transmission, energy, and incorrect inferences. 
    The relative cost reduction for different values of $\beta$, is in the range $14-49\%$. %For instance, the relative cost reduction improvement for $\beta=0.3$ would be equal to $15.22\%$, while for $\beta=0.6$ it would $39.06\%$. Moreover, for a choice of $\beta=0.9$, the cost reduction would be up to $47.36\%$.
%After generating the local inference for a sample, it will be compared to a predefined threshold. If the probability $p$ is greater than the threshold, the inference is considered valid. If not, it will be rejected and offloaded.  
%To calculate the costs, we assume that the cost of offloading (including transmission and delay) is $\beta \in [0,1)$, and the cost of an incorrect inference is $1$. We also make the assumption that the Large ML model is perfect, and therefore any image that is offloaded will be accurately classified. This assumption is only made to enable us to make a comparison with respect to the ideal accuracy. In fact, it is not an unrealistic assumption to consider a perfect large machine learning model, as there are multiple CNNs that have already demonstrated top/exceptional state-of-the-art accuracy on the CIFAR-$10$ dataset, with some models achieving up to $96.612\%$. \\
\begin{table}[ht]
\resizebox{\textwidth/2}{!}{
\begin{tabular}{|l|c|c|c|}
\hline
\multicolumn{1}{|c|}{Approach}      & No Offload      & Full Offload   & Hierarchical Inference \\ \hline \hline 
 Offloaded Images $(\%)$     & $0 (0\%)$       & $10000 (100\%)$ & $3550 (35.5\%)  $      \\ \hline
 Misclassified Images $(\%)$ & $3,742 (37.42\%)$ & $500 (5\%)$        & $1,577 \,  \textbf{ON ED}+ 71 \, \textbf{ON ES} \; (16.48\%)$      \\ \hline
\multicolumn{1}{|c|}{Accuracy $(\%)$}   & $62.58\% $         & $95\% $ & $ 83.52 \% $   \\ \hline
\multicolumn{1}{|c|}{Cost $(\beta)$}   & $3742 $         & $10000 \, \beta + 500 $ & $ 3550 \, \beta + 1648 $   \\ \hline
\end{tabular}}
\caption{Image classification performance comparison for CIFAR-10 dataset: number of offloaded images, misclassified ones and the total cost.} 
\label{table:comparisonCifar}
\end{table}

%% file: dogbreed.tex
\section{HI for Dog Breed Image Classification}

In this section, we introduce another use case that comes in alignment with the concept of  hierarchical inference naturally. Suppose that we have the same set-up (ED connected to an ES) and that we are only interested in classifying the dog breeds of the dogs present in CIFAR-$10$. On one hand, achieving high accuracy for this complicated task using tinyML models is still challenging due to the complex nature of dog breed classification and the limited resources of tinyML devices. On the other hand, fully offloading all of the images can result in high accuracy but is, however, very costly. Under HI, we propose to use an S-ML to classify dog images and non-dog images and only offload the dog images to ES for an L-ML to classify the dog breed. Note that, in this use case the dog images are complex as they require further assistance for classification at the ES. All non-dog images are simple. %that are relevant to our task and thus considered complex data samples (features a dog) and offload them so that the ES classify the dog breeds and disregard the irrelevant images as they are considered simple data samples (does not feature a dog). 

The approach used in this use case generalizes to contexts where the data samples of interest would be sufficiently complex that almost none of them could be accurately inferred on the edge device. Then, the S-ML will not do any final inference for the data of interest. Instead, its job would be just to eliminate a large part of the irrelevant data and only delegate the data samples of interest.
%In our example, all of the images featuring dogs are considered \textit{complex dataset samples}, while the rest are regarded as \textit{simple dataset samples}. 
Table \ref{tab:details} summarizes the performance measures and the current use case details vs the previous ones. 
% \usepackage{multirow}
\begin{table*}[t]
\resizebox{\textwidth}{!}{
\begin{tabular}{|ll|l|l|l|}
\hline
\multicolumn{2}{|l|}{Use Case}                                                & \multicolumn{1}{c|}{Rolling Element Fault Diagnosis}                                 & \multicolumn{1}{c|}{CIFAR-10 Image Classification}             & \multicolumn{1}{c|}{Dog Breed Classification}        \\ \hline
\multicolumn{1}{|c|}{\multirow{2}{*}{Data Description}} & Simple Data Sample  & REB that are in normal state (normal functioning)                                     & Images that would be correctly classified by SML      & Images that are irrelevant, do not feature a dog           \\ \cline{2-5} 
\multicolumn{1}{|c|}{}                                  & Complex Data Sample & REB that are in any faulty state                                                     & Images that would be incorrectly classified by SML    & Images that are of interest, feature a dog                 \\ \hline
\multicolumn{1}{|l|}{\multirow{2}{*}{Cost}}             & Transmission Cost   & -                                                                            & $\beta$ per offloaded image                              & $\beta$ per offloaded image                                   \\ \cline{2-5} 
\multicolumn{1}{|l|}{}                                  & Inference Cost      & -                                                                           & 1 per incorrect inference                             & 1 per offloaded irrelevant image                           \\ \hline
\multicolumn{2}{|l|}{Accuracy}                                                & Number of detected REB in faulty state / Total number of REB in faulty state & Number of correct inferences / Total number of images & Number of correctly classified dogs / total number of dogs \\ \hline
\end{tabular}}
\caption{Data description, cost and accuracy  for the presented use cases.}
\label{tab:details}
\end{table*}
\begin{itemize}
    \item \textbf{S-ML:} The ED will embed a small binary ML model, where it must be able to recognize whether the image features a dog (and thus of our interest) or not. For the S-ML we train a small-size CNN for binary classification, i.e., dog or not dog. The model consists of five layers: $2D$ convolutional layer, MaxPooling2D layer, flatten layer, and two dense layers. The initial dense layer consists of $32$ neurons, utilizing the ReLU activation function. The final layer is a single neuron using the sigmoid activation function. The S-ML produces $p$, a probability score between $0$ and $1$, indicating the probability of the input image being classified as a positive class, which in this scenario is dogs. To make it adequate for resource-constrained devices, the model is afterwards converted to a quantized TFLite model. The resultant model's size is $0.23$ MB and has $63.86 \%$ accuracy.\\ 
\item  \textbf{L-ML:} For this use case, we assume that the L-ML has $100$ \% accuracy, and therefore any image that is offloaded will be accurately classified. Once the S-ML classifies the image as a dog image, the ED will offload it to the ES and the correct dog breed will be determined. This assumption is made since the true labels of the dog breeds are not available for CIFAR-$10$ and thus it is necessary to enable us to make a comparison with respect to the ideal accuracy (ES). In fact, it is not an unrealistic assumption to consider a perfect L-ML, as there are multiple models that have already demonstrated exceptional state-of-the-art accuracy in dog breed classification for various datasets \cite{dogClassifyModel}.
\end{itemize}
For data sample $i$, the S-ML outputs $p_i$, the probability that it features a dog. The ED will utilize this probability to decide whether to offload the sample or consider it irrelevant and disregard it. The ED's objective is to offload only the dog images, and since the SML is a binary classification model, $p = 0.5$ serves as a threshold for dog classification and thus the decision rule for data sample $i$, denoted by $\delta (i)$, will be the following:
\[
    \delta(i) = \left\{\begin{array}{lr}
        \text{Offload} & \text{if } p_i \geq 0.5  \\
        \text{Do not offload} & \text{if }  p_i < 0.5 
        \end{array}\right. 
  \] 
  If the data sample $i$ has to be offloaded to the ES, the cost received will depend on its L-ML's inference. A fixed cost $0 \leq \beta < 1$ will be received for offloading a sample from the dataset of interest (if it was true dog image). Alternatively, if the data sample offloaded was irrelevant (does not feature a dog), a cost of 1 will be incurred.
  
\begin{figure} [ht]
     \centering
     \begin{subfigure}[b]{0.23\textwidth}
         \centering
         \includegraphics[width=\textwidth]{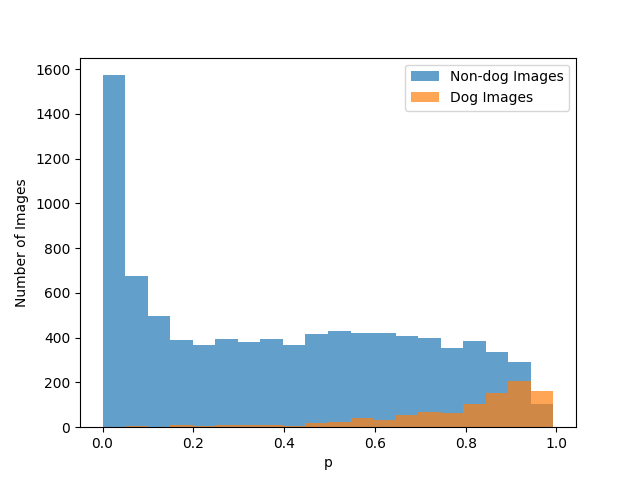}
         \caption{Non-dog images (simple data samples)  and Dog images (complex data samples) of CIFAR-10 dataset as a function of S-ML's output $p$.} \label{dogNondog}
     \end{subfigure}
     \hfill
     \begin{subfigure}[b]{0.23\textwidth}
         \centering
         \includegraphics[width=\textwidth]{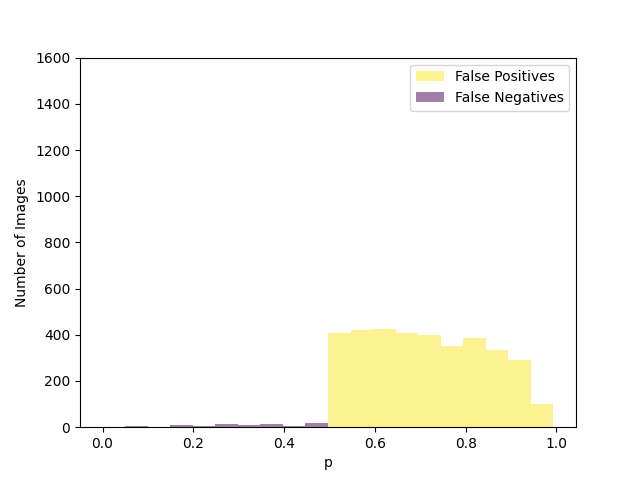}
         \caption{False Negatives (miss-identified dogs) and False Positives (incorrectly classified as dogs) of CIFAR-$10$ by the S-ML.} \label{negatives}
     \end{subfigure}
      \caption{Analysis of our S-ML output for dog breed classification.}
        \label{CIFAR-$10$ Dataset}
\end{figure}

  Figure \ref{dogNondog} represents the actual non-dog (simple data samples) and dog images (complex data samples) in our dataset distributed as a function of $p$. The images having $p \geq 0.5$ will be classified as dog images (complex) by the S-ML. The figure clearly shows that most of the dog images were correctly identified by the S-ML, i.e., the precision of the trained S-ML is high. Specifically, only $88$ dog images out of $1000$ dogs are misclassified and are not offloaded. These are the false negatives. This means that for the task of classifying the dog breeds of CIFAR-$10$, $912$ out of $1000$ dogs will be identified by the S-ML and delegated to the L-ML to determine their correct breed. This yields a $91.2\%$ accuracy. 
  Moreover, the figure illustrates that the S-ML will incorrectly classify a substantial number of non-dog images, $3521$ to be precise, as dog images, leading to their unnecessary offloading to the L-ML. 
  These misclassifications are false positives. 
  %As a matter of fact, there are $3521$ false positives. 
  %This translates to having $3521$ irrelevant data offloaded. 
  An important note here is that these false positives will only affect the total cost incurred but not the accuracy of HI. Figure \ref{negatives} presents the instances of false positives and false negatives as a function of $p$. 

Using the full offload approach for such a scenario can yield maximal accuracy, but it comes with a significant cost of $9000\beta$ for offloading irrelevant images. On the other hand, implementing HI can tremendously reduce the accrued cost while maintaining a reliable level of accuracy. Table \ref{table:dog_HI_cifar10} compares HI with the full-offload approach. Using HI leads to an offload reduction by $55.67 \%$ which reduces the cost by $(\frac{88 \beta + 5479}{10000 \beta + 9000} \cross 100 ) \%$ while maintaining an  accuracy of $91.2\%$.
% Recall that $\beta$ is a design parameter that reflects the designer's preference for balancing the costs of transmission, energy, and number of irrelevant images offloaded. 
 The relative cost reduction for different values of $\beta$, is in the range $50-60\%$.

%HI approach is highly prospective and holds the potential to yield promising and favorable results in various contexts.
From the above use case analysis, we conclude that HI offers a balanced solution that strikes a middle ground between the two extremes of no offload and full offload. On one end, fully offloading the samples can yield maximum accuracy but comes with high costs. On the other end, performing all inferences locally can reduce costs but requires significant compromises in accuracy. HI approach has the potential to offer the best of both worlds, allowing for a significant reduction in costs while still maintaining a reliable level of accuracy.

\begin{comment}
\begin{figure}[ht]%
    \centering
    \subfloat[\centering CIFAR-10 dataset - Non-dog images (simple data samples)  and Dog images (complex data samples) as a function of S-ML's output $p$.]{{\includegraphics[width=5cm]{Uploads/new.png} }} \label{dogNondog}%
    \subfloat[\centering False Negatives (miss-identified dogs) and False Positives (incorrectly classified as dogs) of CIFAR-$10$ by the S-ML]{{\includegraphics[width=5cm]{Uploads/negatives_sameY.png} }} \label{negatives}
    \caption{2 Figures side by side}%
    \label{fig:example}%
\end{figure}
\end{comment}
\begin{comment}
\begin{figure} [ht] 
    \begin{center}
  \includegraphics[width=9 cm]{Uploads/new.png}
    \end{center} \caption{CIFAR-10 simple and complex dataset as a function of the S-ML's output p} \label{dogNondog}
\end{figure} 

\begin{figure} [ht] 
    \begin{center}
  \includegraphics[width=9 cm]{Uploads/negatives.png}
    \end{center} \caption{False Negatives (miss-identified dogs) and False Positives (incorrectly classified as dogs) of CIFAR-$10$ by the S-ML} \label{negatives}
\end{figure} 
%Binary classification of CIFAR-10 dataset to differentiate between images of dogs and other images.
\end{comment}
\begin{table}[ht]
\resizebox{\textwidth/2}{!}{
\begin{tabular}{|l|l|l|}
\hline
Approach                  & Full offload     & HI              \\ \hline \hline 
Number of Offloaded Images & $10,000$                & $4,433$           \\ \hline
Accuracy $(\%)$               & $100\%$              & $91.2\%$            \\ \hline
Cost                   & $1000 \, \beta + 9000$ & $912 \, \beta + 3521$ \\ \hline
Cost Reduction $(\%)$  & $0\%$ & $(\frac{88 \beta + 5479}{1000 \beta + 9000} \cross 100 ) \%$ \\ \hline

\end{tabular}}
\caption{HI binary classification for CIFAR-10 dataset performance comparison with Full offload.}
\label{table:dog_HI_cifar10}
\end{table}  
\begin{comment}
\subsection{Imagewoof / Imagenette - MobileNet 0.25}
Model size is 0.47 MB 
\begin{table}[ht]
\begin{tabular}{|l|l|l|l|l|l|l|}
\hline
Threshold         & 0.1   & 0.15  & 0.2   & 0.3   & 0.35  & 0.4   \\ \hline
Accuracy \%       & 81.59 & 86.44 & 90.21 & 94.47 & 95.72 & 96.73 \\ \hline
Cost Reduction \% & 83.71 & 64.78 & 49.29 & 28.13 & 19.83 & 13.03 \\ \hline
\end{tabular}
\caption{Accuracy and Cost Reduction for classifying Imagewoof dataset using MobileNet, diff thresholds, 10-classes model}
\end{table}

\subsection{Cifar10 - CNN trained }
Model used earlier 

\begin{table}[ht]
\begin{tabular}{|l|l|l|l|l|l|l|l|}
\hline
Threshold         & 0.3   & 0.35  & 0.4   & 0.6  & 0.7   & 0.8   & 0.85  \\ \hline
Accuracy \%       & 66.6  & 68.7  & 70.7  & 86   & 92.1  & 95.1  & 97    \\ \hline
Cost Reduction \% & 89.77 & 86.77 & 82.93 & 59.8 & 48.85 & 38.63 & 33.03 \\ \hline
\end{tabular}
\caption{Accuracy and Cost Reduction for classifying Cifar dataset using SML defined in section 3, diff thresholds, 10-classes model}
\end{table}
\end{comment}

%% file: comparison.tex
%%%%%%%%%%%%%%%%%%%%%%%%%%
\section{Comparison with Existing Approaches}
\begin{figure*} [t]
     \centering
     \begin{subfigure}[b]{0.33\textwidth}
         \centering
         \includegraphics[width=\textwidth]{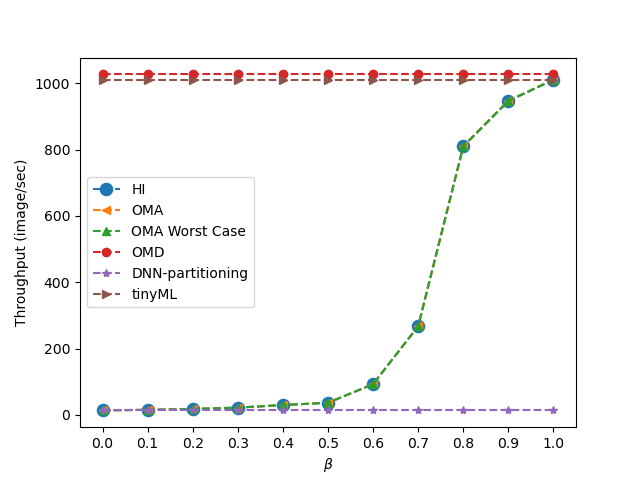}
         \caption{Throughput of CIFAR-$10$ Image Classification\\ of different approaches vs $\beta$.} 
         \label{fig:compare_throughput}
     \end{subfigure}
     \hfill
     \begin{subfigure}[b]{0.33\textwidth}
         \centering
         \includegraphics[width=\textwidth]{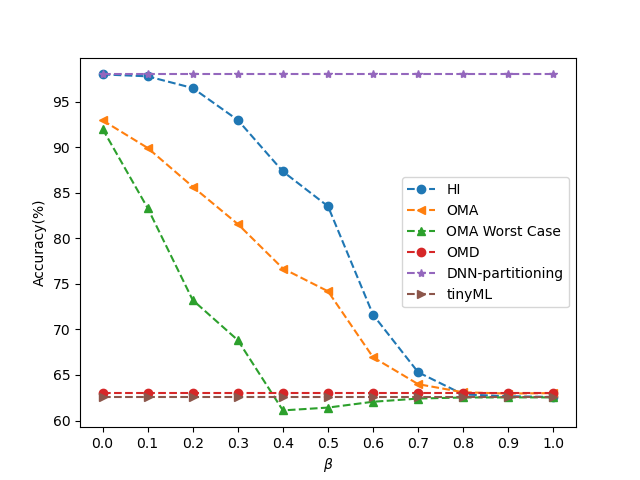}
         \caption{Accuracy of CIFAR-$10$ Image Classification\\
         of different approaches vs $\beta$.}
         \label{fig:compare_acccuracy}
     \end{subfigure}
     \hfill
         \begin{subfigure}[b]{0.33\textwidth}
         \centering
         \includegraphics[width=\textwidth]{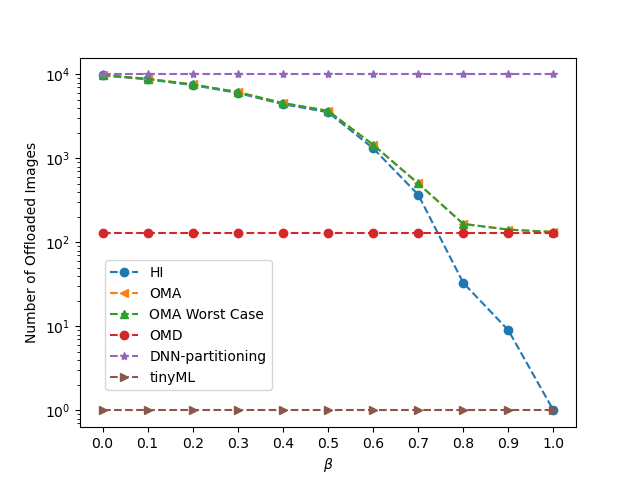}
         \caption{Number of offloaded Jobs to ES of CIFAR-$10$ Image Classification of different approaches vs $\beta$.}
         \label{fig:averagetimejob}
     \end{subfigure}
        \caption{ Different approaches comparison for varying $\beta$.}
        \label{fig:three graphs}
\end{figure*}

This section presents a comparison of HI with tinyML, two algorithms related to inference offloading, and DNN partitioning. We choose the task of CIFAR-$10$ dataset  image classification and throughput, accuracy and number of offloaded images as metrics for our comparison. We begin by introducing these algorithms and describing the experimental setup, followed by a discussion of the results. 
\begin{enumerate}
    \item \textit{TinyML}: Embedded ML approach where all the inferences of the S-ML are accepted (no offload).
    \item \textit{Offloading for maximizing accuracy (OMA)}: computation offloading approach that partitions inference jobs between the ED and the ES for maximizing accuracy given a time constraint. The time constraint we subject is equal to makespan of the HI approach. Two cases are investigated:
    1) Partitioning the images between ED and ES randomly, and 2) \textit{OMA Worst Case} scenario, where the algorithm accepts the local inference for complex samples and offloads the simple ones.
    \item \textit{Offloading for Minimizing Delay (OMD)}: computation offloading approach that partitions the set of images such that S-ML and L-ML will have the same makespan resulting in the minimizing the total makespan for processing all images. 
    \item \textit{DNN-partitioning \cite{Neurosurgeon2017}}:  Since the CIFAR-$10$ dataset has $32 \times 32$ images, the DNN-partitioning approach results in full offload. %The efficient splitting decision with the goal of minimizing delay in the inference and the energy consumption was to have no split and offload all of the images to the ES where the whole DNN (L-ML) is hosted. 
    Further explanation and measurements of why this approach in our context is equivalent to full offload are provided in the Appendix.
\end{enumerate}

We use the S-ML and L-ML described in Section \ref{usecase2} for all the above approaches. Our S-ML can fit on any IoT device but for the current work, we deploy it on a Raspberry Pi $4B$, which  features a single CPU with $4$ cores, a frequency of $1.5 \, GHz$, and $5 \,GB$ of RAM. As for the L-ML, it is hosted on an ES featuring $2$ CPUs with $16$ cores each, a frequency of $2.4 \, GHz$, an NVIDIA Tesla T4 GPU and $256 \, GB$ of RAM. Both devices are connected to the same WLAN and are linked via $802.11$ operating on a $5\, GHz$ channel approved by ETSI for Multi-access Edge Computing (MEC) communication. The transmission of images from ED to ES is accomplished through TCP sockets. In order to estimate the communication time, we utilize iPerf to determine the communication bandwidth. We conduct a total of $30$ experiments, each lasting $60$ seconds, to derive our estimates. The average communication bandwidth between Raspberry Pi and ES recorded was $10.45 \, MB/s$ (SD = 0.6 $MB/s$). The average inference time per one sample by the S-ML on the Raspberry Pi is $0.99 \; msec$, which is significantly lower than the $74.34 \; msec$ required to offload the image from the ED to the ES, and to execute the inference on the L-ML on ES with GPU processing.

We compare HI with the various approaches in terms of throughput (relating to the makespan and thus delay), accuracy and the number of offloaded images (which relates to the approach's cost). Figure \ref{fig:compare_throughput} (\ref{fig:compare_acccuracy}) shows the throughput (accuracy) of the four approaches vs HI for different values of $\beta$. Note that $\beta$ has an effect only on the results of HI, but since the chosen time constraint of OMA (and OMA worst case) is equal to the makespan of HI, we see that OMA (and thus OMA worst case) is also sensitive to the choice of $\beta$. Figure  \ref{fig:compare_throughput}  shows that tinyML has maximum throughput and OMD has throughput close to tinyML throughput. In addition, tinyML scores the lowest number of offloaded images (lowest cost) but Figure \ref{fig:compare_acccuracy} clearly states that their accuracy score is the lowest and not reliable for most of the DL inference tasks. On the other hand, DNN-partitioning has the top accuracy score but it is the worst in terms of throughput and the number of offloaded images. HI and OMA (OMA worst case) have, by design, similar throughput but HI provides significant gains in terms of accuracy and lower number of offloaded images (thus cost) compared to OMA. 
%The results indicate that HI can fully leverage its benefits in scenarios where there is an arrival process for the input. This is because the completion time of jobs is significantly reduced compared to other approaches. In fact, the 

Compared to other approaches, HI offers a significant reduction in latency and cost while maintaining high accuracy. In fact, HI approach was found for example to reduce latency and the number of offload images by approximately $63.15\%$ and $64.45\%$ respectively compared to the full offload approach, while still maintaining $83.52\%$ accuracy for $\beta=0.5$. In summary, HI has demonstrated its ability to provide the best of all worlds, including accuracy, delay, and cost. As a result, it represents a promising solution for real-time applications on resource-constrained devices.

%% file: conclusion.tex
\section{Conclusion}
In conclusion, this paper explored a novel approach, Hierarchical Inference (HI), for performing distributed DL inference between edge devices and edge servers. HI addresses the challenges of resource-constrained EDs in making intelligent decisions using DL inference. Through application use cases and quantitative analysis, we demonstrated the feasibility and benefits of HI. We presented three different use cases for HI:  machine fault detection, CIFAR-$10$ dataset image classification and dog breeds classification. We provide a quantitative comparison between HI and existing techniques for DL inference at the edge which include, tinyML, inference offloading and DNN partitioning. Our results demonstrate that HI achieves significant cost savings while maintaining reliable accuracy. For instance, following HI's approach for CIFAR-$10$ image classification reduces latency and the number of offloaded images compared to the full-offload approach, on average for values of $\beta$ ranging from $0$ to $1$, by $60.84\%$ and $62.179\%$ while maintaining an accuracy greater than $80\%$. With minimal compromise on accuracy, HI achieves low latency, bandwidth savings, and energy savings, making it a promising solution for edge AI systems. 

%% file: appendix.tex
\section{APPENDIX} \label{appendix}
\begin{table}[ht]
\resizebox{\textwidth/2}{!}{
    \centering
    \begin{tabular}{|c|c|c|c|c|c|c|c|c|}
         \hline 
         Device & Processing Unit & L1 & L2 & L3 & L4 & L5 & L6 & L7\\
         \hline \hline
         Raspberry Pi & CPU & 328.9 & 1640.7 & 1131.7 & 970 & 1561 & 1981 & 539.8\\
         \hline
         ES & GPU & 1.01 & 2.51 & 1.50 & 2.16 & 2.31 & 2.89 & 0.91\\
         \hline
    \end{tabular}
    }
    \caption{Time to process each layer of Efficentnet in milliseconds.}
    \label{tab:processing_layers}
\end{table}
\begin{table}[!ht]
    \begin{tabular}{|c|c|c|} 
    \hline
    Input & Size (MB)& Time (msec)\\ 
    \hline\hline
    Image & $0.003$& [$0.28$,$0.30$] \\
    \hline
    Output L1 & $3.06$ & [$276.92$, $310.65$]\\
    \hline
    Output L2 & $1.64$ & [$148.41$, $166.49$]\\
    \hline
    Output L3 & $1.13$ & [$102.26$, $114.72$]\\
    \hline
    Output L4 & $0.97$ & [$87.78$, $98.47$]\\
    \hline
    Output L5 & $1.56$ & [$141.17$, $158.37$]\\
    \hline
    Output L6 & $1.98$ & [$179.18$, $201.0$]\\
    \hline
    Output L7 & $0.53$ & [$47.96$, $53.80$]\\
    \hline
    \end{tabular}
    \caption{Communication time between Raspberry Pi and ES measured in milliseconds.}
    \label{tab:communication_time}
\end{table}

\begin{table}[]
\begin{tabular}{|ll|c|}
\hline
\multicolumn{2}{|l|}{Approach}                                    & Time (msec)         \\ \hline \hline
\multicolumn{2}{|l|}{No Offload}                                  & 0.99                \\ \hline
\multicolumn{2}{|l|}{Full Offload}                                & 74.34               \\ \hline
\multicolumn{1}{|l|}{\multirow{7}{*}{DNN-partitioning}} & Layer 1 & [618.1 , 651.83]    \\ \cline{2-3} 
\multicolumn{1}{|l|}{}                                  & Layer 2 & [2127.78 , 2145.86] \\ \cline{2-3} 
\multicolumn{1}{|l|}{}                                  & Layer 3 & [3211.83 , 3223.57] \\ \cline{2-3} 
\multicolumn{1}{|l|}{}                                  & Layer 4 & [4165.19 , 4175.88] \\ \cline{2-3} 
\multicolumn{1}{|l|}{}                                  & Layer 5 & [5787.27 , 5804.47] \\ \cline{2-3} 
\multicolumn{1}{|l|}{}                                  & Layer 6 & [7803.39 , 7825.21] \\ \cline{2-3} 
\multicolumn{1}{|l|}{}                                  & Layer 7 & [8211.06 , 8216.9]  \\ \hline
\end{tabular}
\caption{Time (in milliseconds) for one inference following different approaches: No Offload, Full Offload and DNN-partitioning at different layers}
\label{tab:comparetimes}
\end{table}
This section provides an explanation of why DNN-partitioning is considered equivalent to full offload approach in our context.
We started by attempting to compare the performance of DNN partitioning when EfficientNet is divided between an Edge Device and an Edge Server. However, after measuring the processing time (as reported in Table \ref{tab:processing_layers}) and the communication time for offloading the features (as reported in Table \ref{tab:communication_time}), we found that it is not feasible to partition the DNN model. Additionally, a single inference on the Edge Device for L-ML takes approximately 8 seconds to execute on the Raspberry Pi, based on the measurements in Table \ref{tab:processing_layers}. 
The inference time required for different approaches, i.e., no offloading, full offloading, and DNN-partitioning at each layers, is presented in Table \ref{tab:comparetimes}. The results indicate that even partitioning at a single layer and deploying a fraction of L-ML at the ED can result in a substantial delay compared to offloading the complete image. For instance, partitioning at a single layer increases the inference time from $74.34$ msec to $[618.1, 651.83]$ msec. This proves that DNN-partitioning is not valid in our case and would lead to a full offload approach. It's worth noting that the output size of each layer in MB is larger than the size of a CIFAR-10 $32 \times 32$ image. This is because EfficientNet requires a $224 \times 224 \times 3$ input, so we up-scaled the CIFAR-10 images using padding. This padding doesn't affect the semantic content of the image, but it does allow them to be scheduled on EfficientNet. Recall that we chose to use EfficientNet because it's the state-of-the-art model for the classification problem with the CIFAR-10 dataset.
% DNN: latency minimizination/ energy minimization. Full offload, local inf, dnn partitioning.